\documentclass[acmsmall, screen]{acmart}

\AtBeginDocument{%
  \providecommand\BibTeX{{%
    \normalfont B\kern-0.5em{\scshape i\kern-0.25em b}\kern-0.8em\TeX}}}

\setcopyright{acmcopyright}
\copyrightyear{2023}
\acmYear{2023}
\acmDOI{}

\acmConference[Ubicomp '23, GenAI4PC Symposium]{Ubicomp '23: ACM Ubicomp}{Oct 8 -- Oct 12, 2024}{Cancún, Mexico}
\acmBooktitle{Ubicomp '23: GenAI4PC Symposium,
  Oct 8 -- Oct 12, 2023, Cancún, Mexico}
\acmISBN{}

\begin{document}

\title{Revolutionizing Mental Health Support: An Innovative Affective Mobile Framework for Dynamic, Proactive, and Context-Adaptive Conversational Agents}

\author{Rahul Islam}
\authornote{Both authors contributed equally to this research.}
\email{mislam5@stevens.edu}
\orcid{0000-0003-3601-0078}
\author{Sang Won Bae}
\authornotemark[1]
\orcid{0000-0002-2047-1358}
\email{sbae4@stevens.edu}
\affiliation{%
  \institution{Charles V. Schaefer, Jr. School of Engineering and Science, Stevens Institute of Technology}
  \streetaddress{1 Castle Point Terrace}
  \city{Hoboken}
  \state{New Jersey}
  \country{USA}
  \postcode{07030}
}








\renewcommand{\shortauthors}{Islam and Bae}

\begin{abstract}
As we build towards developing interactive systems that can recognize human emotional states and
respond to individual needs more intuitively and empathetically in more personalized and context-aware computing time. This is especially important regarding mental health support, with a rising
need for immediate, non-intrusive help tailored to each individual. Individual mental health and
the complex nature of human emotions call for novel approaches beyond conventional proactive and
reactive-based chatbot approaches. In this position paper, we will explore how to create Chatbots that can
sense, interpret, and intervene in emotional signals by combining real-time facial expression analysis,
physiological signal interpretation, and language models. This is achieved by incorporating facial affect
detection into existing practical and ubiquitous passive sensing contexts, thus empowering them with
the capabilities to the ubiquity of sensing behavioral primitives to recognize, interpret, and respond to
human emotions. In parallel, the system employs cognitive-behavioral therapy tools such as cognitive
reframing and mood journals, leveraging the therapeutic intervention potential of Chatbots in mental
health contexts. Finally, we propose a project to build a system that enhances the emotional understanding of Chatbots to engage users in chat-based intervention, thereby helping manage their mood.
\end{abstract}


\received{27 August 2023}
\received[accepted]{1 September 2023}

\maketitle

\section{Introduction}
Long-standing efforts have been made in many fields, including psychology, neurology to train computers to comprehend, decipher, and interpret human emotions and affective states. The term "Affective
computing," first used by Rosalind Picard in the middle of the 1990s, refers to a multidisciplinary
discipline that seeks to create systems and tools that can detect, comprehend, and react to emotional
states in humans. Technology’s promise of offering compassionate support in real time is now closer
than ever with the development of artificial intelligence, machine learning, and natural language processing. This becomes especially important in mental health, where there is a greater need for prompt
and successful therapies than qualified human therapists available.
In recent years, wearable technology that can assess physiological signals like heart rate, temperature, and galvanic skin reaction has made tremendous strides, but connecting them with psychological states is still difficult. Concurrently, affective computing has made strides in understanding facial expressions, a critical component of human communication, and has demonstrated promise in identifying various emotional states and even more complex psychological conditions such as depression. This creates an avenue for deploying these psychological states as context for large language models (LLMs) augmenting in favor of creating a system that depends largely on voice and language, which are intrinsically human modes of expression.
There emerges the possibility of chatbots, which are artificially intelligent conversational agents
that are becoming more commonplace due to the widespread use of smartphones and smart speakers.
Chatbots can be an accessible first point of contact for mental health crises, offering around-the-clock
support and reduce the strain on healthcare systems. Although modern chatbot technologies have
successfully delivered cognitive-behavioral therapies (CBT), their comprehension of and capacity to
react to human emotions is still limited. They still need to be contextualized to capture the subtleties of emotional experiences and respond in a way that is both appropriate and empathetic for the situation.
The dichotomy between our computational capabilities’ sophistication and our affective chatbots’
relative simplicity can be visualized as a graph (Figure 1), plotting emotional predictability against
system transparency. Most current systems occupy a space of high practical application but relatively
low emotional predictability. The vision driving our research is to push the boundaries toward high
emotional predictability without sacrificing practical applicability. Such a system would ideally possess the following characteristics:\\
\textbf{• Emotional predictability}: The ability to recognize and interpret various complex emotions the user expresses accurately.\\
\textbf{• Contextual responsiveness}: Being able to respond appropriately according to the emotional state and context of the user.\\
\textbf{• Privacy preservation}: Should be designed with privacy at its core. An ideal system would not compromise the user’s sensitive data, process data on devices, or anonymize any
collected information.\\
\textbf{• Easy accessibility}: The system should be available on commonly used platforms like smartphones and require minimal setup or calibration by the user.\\
\textbf{• Cost-effectiveness}: The system should not require expensive hardware or software, making
it a feasible consumer product in the near future.

We will thus focus on developing affective computing techniques that elevate the emotional
predictability of mental health chatbots while preserving their transparency and accessibility. We will also delve into this technology’s ethical and privacy-related challenges, aiming to ensure these solutions are as responsible as they are innovative.

\begin{figure}[h]
\centering 
    \includegraphics[scale=1]{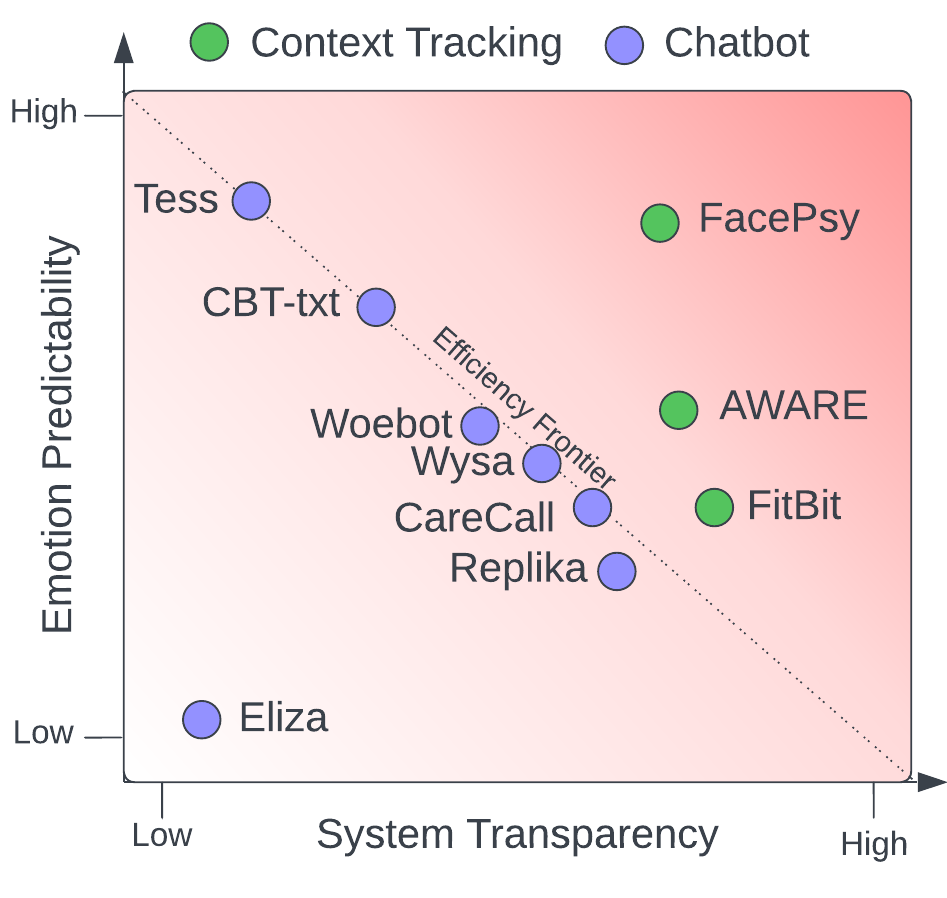}
    \caption{A design space plotting emotional predictability vs. system transparency. Most approaches lie
along the diagonal, where system transparency decreases as emotional predictability increases. Ideally, we want systems
with higher emotional predictability without commensurate system transparency trade-offs. Note: FacePsy refers to the affective mobile framework we have developed.}
    \label{fig:Proposed}
\end{figure}

\section{Building a Context-Adaptive Chatbot
System}
Our goal for this proposal is to discuss an exploration of the feasibility of building a chatbot system that integrates affective computing and language model-based chatbots to monitor user affective states. This includes adapting context, allowing users to converse and understand why the system categorized the user’s input as a particular emotional state for shared decision support to engage user in cognitive behavioral therapy (CBT). The system will comprehend the text and recognize and respond to the user’s emotional context based on valence and arousal. It will analyze facial behavior, head gestures, pupil-iris ratio,
app usage, and physical activity to detect the user’s affective state, creating an interactive agent that provides a more nuanced and emotionally adaptive conversational experience while preserving their transparency and accessibility.

\subsection{Integration of Affective Computing and Intelligent Chatbot For Therapeutic Care}

Harnessing the advancements in affective computing, particularly in the realm of facial expression
recognition, facial action coding systems (e.g., AUs), head gestures, physiological indicators (e.g.,
pupil-iris ratio), digital activity (e.g., app usage pattern), and physical activity (e.g., steps), we plan to incorporate inferences from these capabilities into the chatbot system. This can be done either via integration with the chatbot interface or by developing companion applications that feed affective data to the chatbot in real-time. The key challenge here is to inform the user when the system detects the user’s inaccurate affective states. Improving chatbot ability create an interaction to help users understand how the system works, make informed decisions, judge when to trust the system, and learn how accurate it is.

Building on the affective recognition capabilities, the chatbot will employ CBT tools to facilitate
user engagement and aid in mental health support. Chatbots can offer lightweight therapy or counseling to individuals with milder symptoms of depression, serving as a real-time, first line of defense. The
system will provide tools for cognitive reframing ,\cite{sharma2023cognitive} mood journaling \cite{kawasaki2020assessing}, and gratitude journaling
\cite{lives2021chatbot}, and these tools will be context-aware, changing and adapting according to the detected emotional states. The system will be combined with sensing modules such as smartphone camera, phone sensor, and activity sensor, to monitor facial behavior, head gestures, pupil-iris ratio, app usage, and physical activity, and predict moods. This provides data-driven insights into mental health experiences, such as triggers, physical manifestations of mood, and near-future mood predictions \cite{wolf2020designing}.

While using a camera as a sensor, among other sensors, raises questions about user data and
privacy, we focus on comprehension and sense-making around the underlying AI model. Concerns over
global and local explanations in the context of chatbots are traditional. When communicating with
a human, a chatbot should be able to explain why it categorized the user’s input as a particular
emotional state and give a specific response. As we implement this system, we need to consider how
machine learning aligns with therapeutic experiences. While machine learning relies on predicting historical data, mental health conditions like depression do not always follow a linear progression. Instead, it is situationally emergent. As such, we can explore reflecting dynamism in modeling historical data and explaining (mis)alignments to the humans through XAI technologies.

\subsection{Exploration of Context-Adaptive Chat-Based Intervention System Using XAI}

One possible drawback of utilizing machine learning models is that they tend to make mistakes. Although machine learning models are capable of providing precise emotional state detection to user
context, they may, on occasion, detect user emotional states that are inaccurate. This can result in user frustration and dissatisfaction. We could assess users’ evaluations of chat interface designs for various components involved in predicting mood via a user’s facial behavior, head gesture, pupil-iris ratio, app usage, and physical activity. Specifically, we could aim to measure the importance of the information presented through the chat interaction and the level of agreement with statements related to the interaction’s understandability. We could also aim to evaluate the interaction’s ability to help users understand how the system works, make informed decisions, judge when to trust the system and learn how accurate it is.
To achieve this goal, we can integrate a response decision matrix with a question-driven framework \cite{liao2020questioning} and complement it with a clinical decision support system \cite{schoonderwoerd2021human}. This integration aims to enhance the transparency of the mood prediction model. Specifically, we could utilize the XAI question bank developed by Liao et al. \cite{liao2020questioning} to identify a list of explanation methods supported by XAI algorithms. We could focus on methods that generate post-hoc explanations for opaque machine learning models and interpolate them to a chat interaction context.

\section{Domain Contribution}
\subsection{Context Adaptive Chatbots for Enhanced User Empowerment}
The domain of context-aware pervasive systems, particularly chatbots, greatly benefits from integrating affective computing and Explainable AI (XAI) to build an adaptive chat interaction with the
right therapy tools. Here’s why:
\textbf{(1) Enhanced Personalization and Real-time Feedback}: Smartphones are personal computing devices that accompany users throughout their daily routines. We can offer a highly personalized
chat experience by integrating context awareness into chatbots. For instance, a chatbot that understands a user’s current emotional state can adapt its responses to be more empathetic or cheerful, depending on the situation, providing tools just in time and allowing for real-time feedback. This immediacy can be leveraged to give users instant insights into their emotional states, helping them make informed decisions or better understand their feelings.
\textbf{(2) Privacy-Preserving Context Awareness}: Mobile devices, being personal, can offer a level
of privacy that might not be possible with external systems. For instance, facial recognition for
emotion detection can be done on-device without sending data to external servers, ensuring user data
privacy. Allowing users to control what datapoint machine learning model can use for predicting the
mood hence limiting the sensor use permission by the system for privacy-preserving.
\textbf{(3) Decision Support on-the-go}: Chatbots equipped with XAI can serve as decision support
tools, explaining their suggestions or actions. For instance, if a user is detected as sad, the chatbot
might suggest a walk and help the user break the thinking pattern blocking their action. When the user asks why, the chatbot can explain that physical activity has been shown to boost mood, and it
noticed the user hasn’t been active for a while.
In conclusion, the domain of context-aware pervasive chatbots systems stands to gain a richer,
more personalized, and transparent user experience by integrating affective computing and XAI. This
enhances user trust and engagement and paves the way for chatbots to be more than just transactional
tools – they are companions that understand and adapt to the user’s emotional state.

\section{Building Trust through Transparent AI in Therapeutic Interventions}

\subsection{User Engagement with AI for Managing Depressive Symptoms}

Individuals with depression often experience persistent negative thinking patterns that can exacerbate their symptoms \cite{price2020neuroplasticity}. Recognizing this, AI-driven interventions, particularly chatbots, can play a pivotal role in assisting these individuals. Specifically, AI can facilitate using Cognitive Behavioral Therapy (CBT) tools, such as cognitive restructuring, to help users reframe and challenge their negative thoughts and break the cycle. This is crucial as timely interventions prevent escalating depressive symptoms and promote better mental well-being. Existing conversational agents and AI therapies, like Woebot and Wysa, have already set a precedent in this domain. They offer real-time, interactive support, guiding users through therapeutic exercises and providing instant feedback, thereby playing a significant role in the decision-making processes of individuals with depression.

\subsection{Emotionally Adaptive Conversational Experience}
The emotionally adaptive conversational experience ensures that the chatbot system can assign the user’s emotional state by detecting their pleasure (i.e., valence) and alertness (i.e., arousal) and adapt its interactions based on the assigned state. This involves a dynamic and iterative process where
the system continuously learns from user interactions, updates its learning, and remembers past interactions to provide a more personalized and empathetic user experience. 

\textbf{Step 1: Adaptation of Emotions via Context Information}

\textit{(a) Real-time Sensing}: The system integrates various sensors and data sources, such as facial expression recognition, head gestures, and physiological indicators like pupil-iris ratio. By analyzing this data in real time, the system can infer the user’s current pleasure and alertness levels and assign mood states accordingly.

\textit{(b) Historical Data Analysis}: The system will retain past interactions and emotional states. Analyzing patterns over time with reinforcement learning can identify triggers or recurring emotional patterns,
allowing for more proactive support by delivering a just-in-time adaptive intervention (JITAI).

\textbf{Step 2. Dynamic Updating and Memorization}

\textit{(a) Continuous Learning and Feedback Mechanism}: As users interact with the chatbot, the system continuously updates its understanding of their emotional patterns. This is achieved through machine learning techniques that allow the model to refine its predictions based on new data.

\textit{(b) Long-term Memory in Conversation}: The system will have a memory component that remembers past interactions, emotional states, and user feedback. This ensures the chatbot can reference past interactions for context, making the conversation more continuous and personalized. Because it is inefficient to use the entire conversation history as long-term memory, we intend to use LLM prompting
techniques for obtaining and managing information from conversation history.

\subsection{Transparency, Trust, and User Support in AI-driven Therapeutic Systems}

System transparency is paramount in AI-driven therapeutic systems because it underpins user trust. Trust is the linchpin for user engagement, especially in sensitive areas like mental health. When users understand how the system works, they are more likely to trust its suggestions and engage in therapeutic conversations. The future AI system should offer real-time feedback and facilitate user
self-reflection and self-regulation. Users can make informed decisions about their mental well-being by presenting algorithmically suggested actions. Furthermore, the future conversational agent should 
leverage LLMs to provide contextually relevant and empathetic responses. By integrating LLMs,
the system can generate more human-like interactions, enhancing the user’s comfort and trust in the
chatbot. As such, the innovative affective mobile system for mental health support aims to seamlessly blend advanced AI techniques with therapeutic tools, ensuring transparency, fostering trust, and providing robust user support.

\bibliographystyle{ACM-Reference-Format}
\bibliography{sample-base}

\end{document}